\documentclass[reprint,amssymb,aps,showkeys,]{revtex4-2}

\usepackage{graphicx}
\usepackage{url}
\usepackage{multirow}
\usepackage{color}
\usepackage{amsmath}

\newcommand{\ds}{\displaystyle}

\begin{document}

\title{Generating quantum feature maps for SVM classifier}

\author{Bang-Shien Chen}
\email[]{dgbshien@gmail.com}
\affiliation{Department of Mathematics, National Taiwan Normal University,
Taipei 11677, Taiwan}

\author{Jann-Long Chern}
\email[]{chern@math.ntnu.edu.tw}
\thanks{The author's work is supported by National Science and Technology Council,
Taiwan (NSTC 110-2115-M-003-019-MY3).}

\affiliation{Department of Mathematics, National Taiwan Normal University,
Taipei 11677, Taiwan}

\date{\today}

\begin{abstract}
    We present and compare two methods of generating
    quantum feature maps for quantum-enhanced support vector machine, a
    classifier based on kernel method, by which we can access high dimensional
    Hilbert space efficiently. The first method is a genetic algorithm with 
    multi-objective fitness function using penalty method, which incorporates 
    maximizing the accuracy of classification and minimizing the gate cost of 
    quantum feature map's circuit. The second method uses variational quantum
    circuit, focusing on how to contruct the ansatz based on unitary matrix
    decomposition. Numerical results and comparisons 
    are presented to demonstrate how the fitness fuction reduces gate cost while
    remaining high accuracy and conducting circuit through unitary matrix obtains
    even better performance. 
    In particular, we propose some thoughts on reducing and optimizing the gate cost
    of a circuit while remaining perfect accuracy.
\end{abstract}

\keywords{support vector machine; quantum feature map; genetic algorithm}

\maketitle

%------------------- Section 1--------------------------

\section{Introduction}\label{sec1}
Quantum computing is a type of calculation using quantum mechanical properties,
such as superposition, interference, and entanglement. It is well known
that quantum computing is capable of solving certain computational problems
faster than classical computers. Applying quantum mechanics on machine learning 
\cite{BWP17,SSP14}
is expected to have a performance on speeding up calculations. As for the
classifier, quantum kernel method via support vector machine \cite{HCT19},
among other development by different research groups
\cite{BYH22,GGC22,LAT21,MBS22,PQW20,RML14,SCH21,SYG20}, has been attested a
powerful mean of using high dimensional quantum state space. These studies show
the possibility of machine learning using quantum computers, which may have a
boost in the future.\\

However, a suitable feature map, corresponding to a suitable kernel, plays a big role
in the kernel method. A quantum circuit is employed to produce a feature map and takes
the dataset from its original low dimensional real space to high dimensional
quantum state space, i.e., the Hilbert space \cite{SK19}. In general, the quantum circuit should be
designed to increase its capacity to explore the Hilbert space and encode
probability distribution more efficient but also avoid expensive gate cost and
circuit size to decrease quantum noise. Taking this into account, we first
modified the method proposed in \cite{ARG21},
which automatically generates quantum feature maps by 
using multi-objetive genetic algorithm \cite{CYH16,LIN16} both to maximize
the accuracy and to minimize the size of the circuits. Then, we proposed
another method learning through parameterized variational circuit \cite{KMT17} with
different ansatz constructed by unitaty decomposition \cite{ICK16,KSA21,SBM06,VW04}. 
\\

The paper is organized as follows. A brief review of support vector machine and
quantum kernel method is presented in Section \ref{sec2}. Section \ref{sec3} describes
the genetic algorithm in more details, including how we encode the problem 
and how we optimize the multi-objective fitness function. Section \ref{sec4} explains
the variational method we proposed, using hardware efficient ansatz and unitary 
decomposition ansatz.
Section \ref{sec5} presents the results of applying the algorithm to
three datasets: moonshape data, ad hoc data and a dataset generated by a specific group
structure. Finally, we draw our conclusions
and summarize our learned lessons in Section \ref{sec6}.

%------------------- Section 2--------------------------

\section{Quantum Kernels}\label{sec2}

\subsection{Support vector machine}\label{sec2-1}
Support vector machine (SVM) is a widely used classical supervised classification model in machine learning.
Suppose we have a set of data points 
$\{(x_i,y_i)\,| \, x_i\in \mathbb{R}^n, \, y_i= \pm1 \}$, SVM finds a
pair of parallel hyperplane $w \cdot x+b= \pm 1 $ that divides the data points
into two classes, where the decision boundary corresponds
to the hyperplane $w \cdot x+b= 0 $ with orientation controlled by $w$ and offset controlled by $b$.
The goal is to differ the data points by classes as much as possible, i.e.,
maximize the margin (distance) between the parallel
hyperplane, where we can express the problem as $\min \frac{1}{2}
\Vert w \Vert_2 ^2 \ \text{s.t.} \ y_i(w \cdot x_i+b) \geq 1$.
The optimization problem for the primal problem can be formulated
into Lagrange dual problem:

\begin{equation} \label{L_dual}
    \begin{array}{cl}
        \max &  L(w,b,\alpha) = \ds \sum_{i=1}^{n}\alpha_i - \frac{1}{2}\sum_{i=1}^{n}\sum_{j=1}^{n}
        \alpha_i\alpha_jy_iy_jx_ix_j \\
        \text{s.t.} & \ds \sum_{i=1}^{n} \alpha_i y_i=0,\ \alpha_i \geq 0
    \end{array}
\end{equation}\medskip

When the data points are not linearly separable, we use a feature map $\phi$ to map
the data into high dimensional space for more features or to low dimensional space
to eliminate unimportant features. The kernel trick is to offer a more efficient
and less expensive way to transform data into higher dimensions, where the data
can be linearly seperated in the higher dimensional space. The kernel
function, denoted as $K(x_i,x_j)=\phi(x_i)\phi(x_j)$, allows us to perform SVM
by just knowing how to calculate the inner product of $\phi(x_i)$ and $\phi(x_j)$
instead of knowing what the feature map $\phi$ is. In particular, the RBF kernel,
i.e., the Gaussian kernel is often used. Applying the kernel trick, we
can rewrite the Lagrange dual problem:

\begin{equation}\label{L_dual_k}
    \begin{array}{cl}
        \max & L(w,b,\alpha) = \ds \sum_{i=1}^{n}\alpha_i - \frac{1}{2}\sum_{i=1}^{n}\sum_{j=1}^{n}
        \alpha_i\alpha_jy_iy_j K(x_i,x_j) \\
        \text{s.t.} & \ds \sum_{i=1}^{n} \alpha_i y_i=0,\ \alpha_i \geq 0
    \end{array}
\end{equation}

\subsection{Quantum kernel method}\label{sec2-2}
A big challenge in quantum machine learning is how we encode the classical information
to quantum states for quantum computing.
The kernel method can be implemented to quantum computing by considering the quantum circuit
as an encoding function $\Phi$ where the quantum feature map is defined
as $\vert \Phi (x) \rangle = U(x) \vert 0 \rangle ^n$
and the kernel is naturally defined as
$K(x_i,x_j)= | \langle \Phi (x_i) \vert \Phi (x_j) \rangle |^2 $. This provides the
data points to be mapped into quantum Hilbert space ~\cite{SK19}, where using quantum computer will
have a big advantage.
Using quantum circuits, a data $x \in \mathbb{R}^n$ is mapped to an n-qubit quantum
feature state $\Phi(x) = U(x)|0^n\rangle\langle 0^n|U^\dagger(x)$ through a unitary
circuit $U(x)$, and we can employ the circuit to evaluate the kernel by: 

\begin{equation}\label{K_circuit}
    K(x_i,x_j) = |\langle\Phi(x_i)|\Phi(x_j)\rangle|^2
    = |\langle 0^n|U^\dagger(x_i)U(x_j)|0^n\rangle|^2
\end{equation}\medskip

The kernel can be estimated on a quantum computer by evolving the initial state
$|0^n\rangle$ with $U^\dagger(x_i)U(x_j)$ and counting the frequency
of the $0^n$ outcome. As long as the qauntum kernel $K(x_i,x_j)$ can be computed effciently,
we can do optimization in the Hilbert space while other optimization steps can be performed
on classical computer.
For more details about quantum kernel estimation,
please refer to ~\cite{HCT19}. Note that the ZZFeatureMap they provided
is an unitary circuit defined as
$U=\widetilde{U}_{\Phi(x)}H^{\otimes2}\widetilde{U}_{\Phi(x)}H^{\otimes2}$, where
$\widetilde{U}_{\Phi(x)}=\exp (i\Phi_1(x)ZI+i\Phi_2(x)IZ+i\Phi_{1,2}(x)ZZ)$, is
only suitable for a specific dataset. The main challenge is to seek suitable kernel
function for different datasets, which led to researches on automatically
learning kernels \cite{ARG21,GA11}.

%-------------------- Section 3---------------------

\section{Genetic Algorithm} \label{sec3}

Genetic Algorithm is a evolutionary algorithm inspired by
the process of natural selection, which has extensively been used to solve
optimization problems. The
evolution starts from a population of randomly generated chromosomes, 
with the population in each iteration called generation.
The fitness of every chromosome in the population is
calculated, selected from the current population, and each chromosome's gene is 
modified to form a new generation for the next iteration of the algorithm.

\subsection{Chromosome and Population}\label{sec3-1}

A chromosome is a potential solution to the problem being solved,
consists of a set of genes/numbers. Note that encoding problems to individual
genes is often the hardest, a careful analysis between the circuit and numbers
is necessary. We modified the encoding method \cite{ARG21} as follows,
the first three bits determine whether the gate is a
Rotation gate, Hadamard gate, CNOT gate or identity gate, where any unitary
operation can be approximated to Rotation gate, Hadamard gate and CNOT gate.
If the first three bits determine to
be a Rotation gate, the last three bits determine a coefficient between $[\pi/8, \pi]$
for the rotation parameter. As for the circuit size, we use 2-qubits in order to
compare with ZZFeatureMap later. The $i$\textsuperscript{th} gate acts on $i$ (mod 2) qubit, CNOT gate
acts on $i$ (mod 2) and $i$+1 (mod 2) specifically. We added more bits to determine
the rotation angle for more precisely results.
The model starts
by generating a binary string with the size $M \times N \times 6$, where $M$ is the
number of chromosomes and $N$ is the number of genes.

\begin{figure}[htbp]
    \centering
    \caption{{Modified encoding example based on \cite{ARG21}.}}
    \label{encode}
    \includegraphics[width=0.48\textwidth]{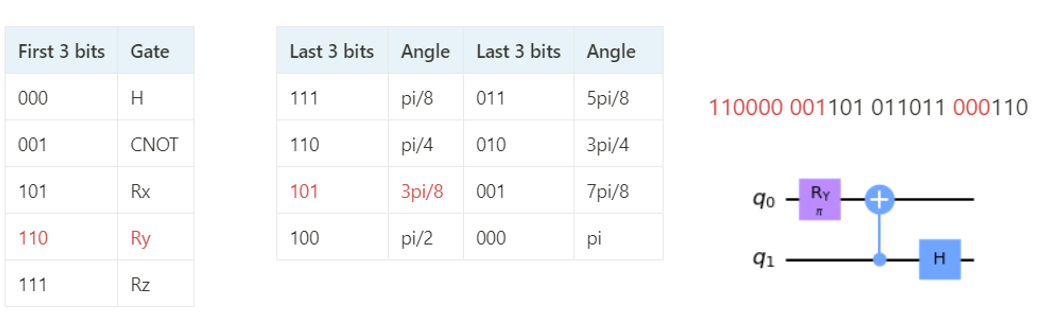}
\end{figure}

\subsection{Fitness and Matting pool}\label{sec3-2}
Fitness is also considered as objective function or cost function. We pass a solution
to this function and it returns the fitness of that solution. In the below multi-objective
fitness function (\ref{FitnessFunc}), we aim to both {\bf maximize the accuracy} and 
{\bf minimize the gate cost}. A common strategy for multi-objective problem is to find the pareto front
\cite{ADB14}
by choosing high domination points with crowd distance techniques, however, pareto front
method does not suit our situation because high-accuracy performance is usually hard to
find. If we don't give more weight to the accuracy factor, the model will be
influenced more by gate cost, thus it tends to find low-cost circuit rather than high-accuracy
circuits. The accuracy is simply the mean accuracy of the given test data and
label, with the QSVM model generated by a quantum circuit. We use the
gate cost proposed in \cite{LLK06}, by counting the sequence of basic physical
operations required for implementation on a quantum computer:

\begin{equation}\label{GateCost}
    \text{GateCost} = \text{Rgate} + 2\text{Hgate} + 5\text{CNOTgate}
\end{equation}

With the two ingredients, we can define the multi-objective fitness function. Our 
main problem is to minimize the gate cost with a constraint of maximizing 
the accuracy. We consinder the reciprocal of square of accuracy
as the penalty function and replace the problem as follows, where $w$ is a 
positive constant (penalty weight) that depends on different datasets.

\begin{equation}\label{FitnessFunc}
    \text{Fitness} = \text{GateCost} + w/\text{Accuracy}^2
\end{equation}

The matting pool chooses the parents for the next generation. The simplest method
is to first determine a pool size and then choose the best solutions based on the
pool size. Note that the same chromosome's fitness should be calculated in every generation,
due to the fact that the accuracy are encoded by probability distribution,
computing in every generation guarantees the stability of the outcome by a chromosome. 
For the mutation operation, we set a probability at $80\%$
to randomize a chromosome and simply do a bit flip.

%------------------ Section 4------------------------

\section{Variational Ansatz} \label{sec4}

In quantum mechanics, every quantum circuit is equivalent to an unitary matrix.
Because that the feature map is calculated by a quantum circuit,
it is also calculated by a specific unitary matrix, which our goal is to 
find the unitary matrix corresponding to a feature map that suits the dataset.
We propose two ways to construct a variational quantum circuit with 
learnable parameters in order to find the unitary matrix.

\subsection{Hardware Efficient Ansatz}\label{sec4-1}
We use the hardware efficient ansatz, which is spanned by special unitaries and
entangle layers, proposed in \cite{KMT17}. This ansatz is often used as the parameterized
circuit for variational quantum eigensolver \cite{PMS14}, consisting of a RY gate layer, 
RZ gate layer, then a CNOT gate layer. One can simply add the depths by repeating
the layers where the ansatz usually ends with the rotaion layer. We encode the
linear combination of the data points as the parameter of each rotation gate.
An one depth 2-qubit feature map can be formulated as follow:

\begin{equation}\label{he_ansatz}
  \begin{array}{c}
      (e^{I\phi_1 Z}\otimes e^{I\phi_2 Z})(e^{I\phi_3 Y}\otimes e^{I\phi_4 Y}) \\
      \\[-0.6em]
      \phi_j = a_j x_1 + b_j x_2\\
  \end{array}
\end{equation}

\begin{figure}[htbp]
  \centering
  \caption{Hardware efficient ansatz.}
  \label{he_circuit}
  \includegraphics[width=0.36\textwidth]{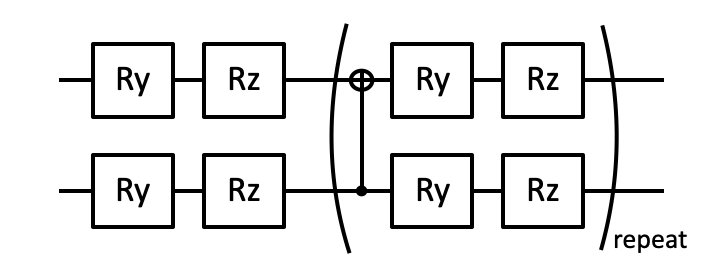}
\end{figure}

After the circuit is constructed, we set the accuracy of the QSVM model as
the cost function and update the parameters through classical optimizer 
COBYLA (constraint optimization by linear approximation). Note that COBYLA 
only performs one objective function evaluation per optimization iteration.

\subsection{Unitary Decomposition Ansatz}\label{sec4-2}
Due to the fact that quantum circuits are equivalent to unitary matrices,
we use the unitary decomposition ansatz proposed in \cite{SBM06} in order to 
produce arbitrary unitaries. Given an even size
unitary matrix $M$, which the matrix we want to encode to a quantum circuit is 
always the size of $2^k$, it can be decomposed to smaller unitaries $L_1,L_2,R_1,R_2$
and diagonal matrices $C,S$ with the size of $2^{k-1}$ 
by Cosine-Sine Decomposition such that $C^2+S^2=I_{2^{k-1}}$, as below:

\begin{equation}\label{cs_decomposition}
  \begin{array}{l}
    M = \begin{bmatrix}L_1&0\\0&L_2\end{bmatrix}\begin{bmatrix}C&-S\\S&C\end{bmatrix}\begin{bmatrix}R_1&0\\0&R_2\end{bmatrix} 
  \end{array}
\end{equation}

The cosine-sine matrix is in the form of multiplexor gate, while the left and right
factor can further be decomposed. Given an unitary matrix $X$ with the form of 
$X_1 \oplus X_2$, it can be decomposed to $(I \otimes V)(D \oplus D^\dag)(I \otimes W)$
where $D^2, V$ are the eigenvalues and eigenvectors of the marix $X_1X_2^\dag$, that is,

\begin{equation}\label{eg_decomposition}
  \begin{array}{l}
    \begin{bmatrix}X_1&0\\0&X_2\end{bmatrix} = \begin{bmatrix}V&0\\0&V\end{bmatrix}\begin{bmatrix}D&0\\0&D^{\dag}\end{bmatrix}\begin{bmatrix}W&0\\0&W\end{bmatrix} 
  \end{array}
\end{equation}

$V,W$ are now smaller unitary matrices and the diagonal matrix is also in the form of 
multiplexor gate. The whole process is called the Quantum Shannon Decomposition,
please refer to \cite{SBM06} for more deatails. At last, the single qubit unitary matrices
can be decomposed by Euler Decomposition (ZYZ Decomposition). Again, We encode the
linear combination of the data points as the parameter of every rotation gates.
A 2-qubit feature map can be formulated as follow:

\begin{equation}\label{ud_ansatz}
  \begin{array}{c}
      U_{\phi_{1,2,3}} M_{\text{R}_{\text{Z}}\phi_{4,5}}
      U_{\phi_{6,7,8}}M_{\text{R}_{\text{Y}}\phi_{9,10}}
      U_{\phi_{11,12,13}} M_{\text{R}_{\text{Z}}\phi_{14,15}}
      U_{\phi_{16,17,18}} \\
      \\[-0.4em]
      U_{\phi_{j,j+1,j+2}} = \exp({i\phi_j \text{IZ}+i\phi_{j+1} \text{IY}+i\phi_{j+2} \text{IZ}})\\
      \\[-0.6em]
      \phi_j = a_j x_1 + b_j x_2
  \end{array}
\end{equation}

\begin{figure}[htbp]
  \centering
  \caption{Unitary decomposition ansatz.}
  \label{ud_circuit}
  \includegraphics[width=0.36\textwidth]{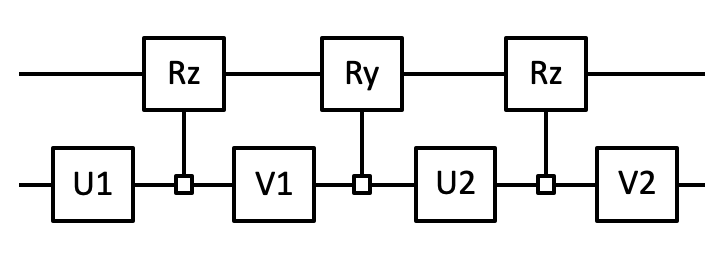}
\end{figure}

\section{Results} \label{sec5}

\subsection{Moonshape dataset}\label{sec5-1}
\begin{figure}[htbp]
    \centering
    \caption{Moonshape data using genetic algorithm.}
    \label{ga_moon}
    \includegraphics[width=0.48\textwidth]{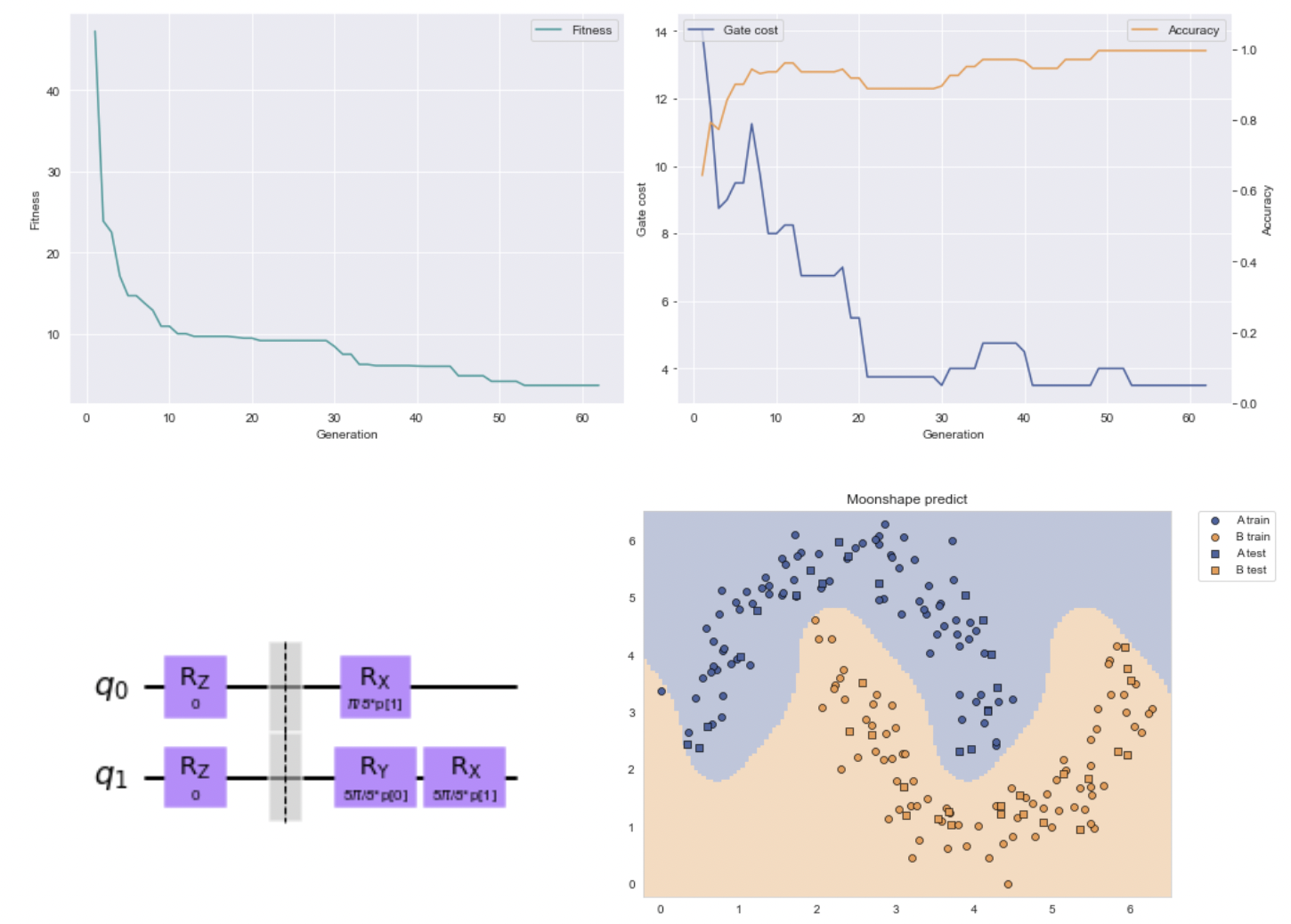}
\end{figure}

First, we consider the moonshape dataset generated by Sklearn \cite{PVG11}.
By using genetic algorithm, we choose the weight $w$ = 20 in equation (\ref{FitnessFunc}) because the data here is rather
simple, which we do not put much emphasis on the accuracy factor.
The genetic algorithm produces the simple uncorrelated circuit with low gate cost
and perfect accuracy as illustrated in Figure \ref{ga_moon}.
If we choose lower weight, the algorithm will produce some circuit with only one
gate cost but a lower accuracy, due to the fact that this particular circuit's
fitness is lower than any other circuits with perfect accuracy. If we choose
higher weight, the algorithm will eventually find the optimal solution but with
more time cost.
Note that the circuit can be optimized if the gates share the
same rotation axis and parameter, and can be combined to a single rotation gate
with additional angles.

By using variational circuits, both hardware efficient ansatz and unitary decomposition
produce perfect accuracy. For hardware efficient ansatz, we choose to use only one depth
such that the circuit's gate cost is only 4, again the data here is rather simple.
A disadvantage of unitary decomposition ansatz is its high gate cost, unlike hardware efficient ansatz,
the circuit is fixed.

\begin{figure}[htbp]
  \centering
  \caption{Moonshape data using hardware efficient ansatz.}
  \label{he_moon}
  \includegraphics[width=0.48\textwidth]{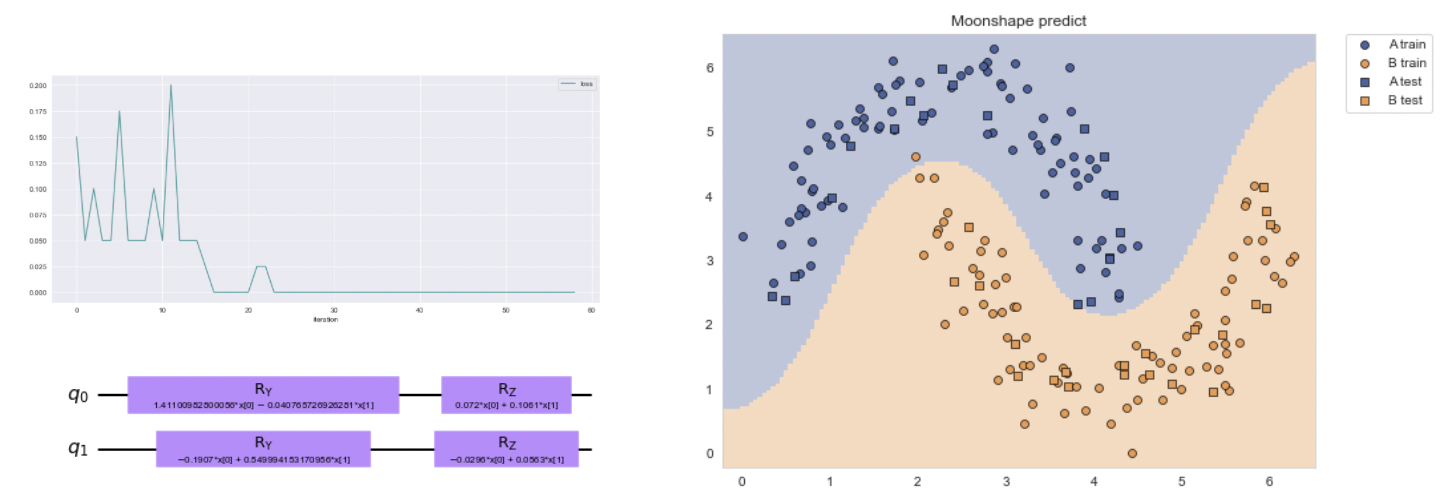}
\end{figure}

\begin{figure}[htbp]
  \centering
  \caption{Moonshape data using unitary decomposition ansatz.}
  \label{ud_moon}
  \includegraphics[width=0.48\textwidth]{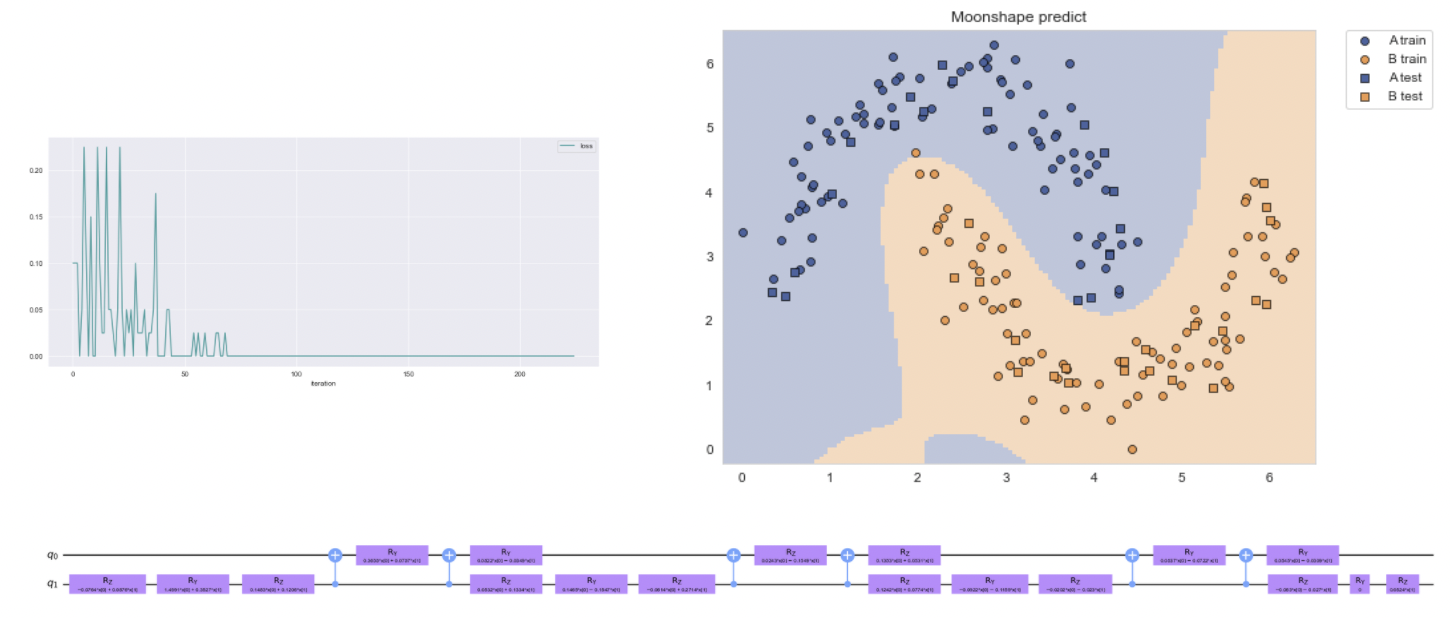}
\end{figure}

\subsection{Adhoc dataset}\label{sec5-2}
\begin{figure}[htbp]
  \centering
  \caption{Ad hoc dataset using genetic algorithm (left) and ZZFeatureMap (right).}
  \label{ga_adhoc}
  \includegraphics[width=0.48\textwidth]{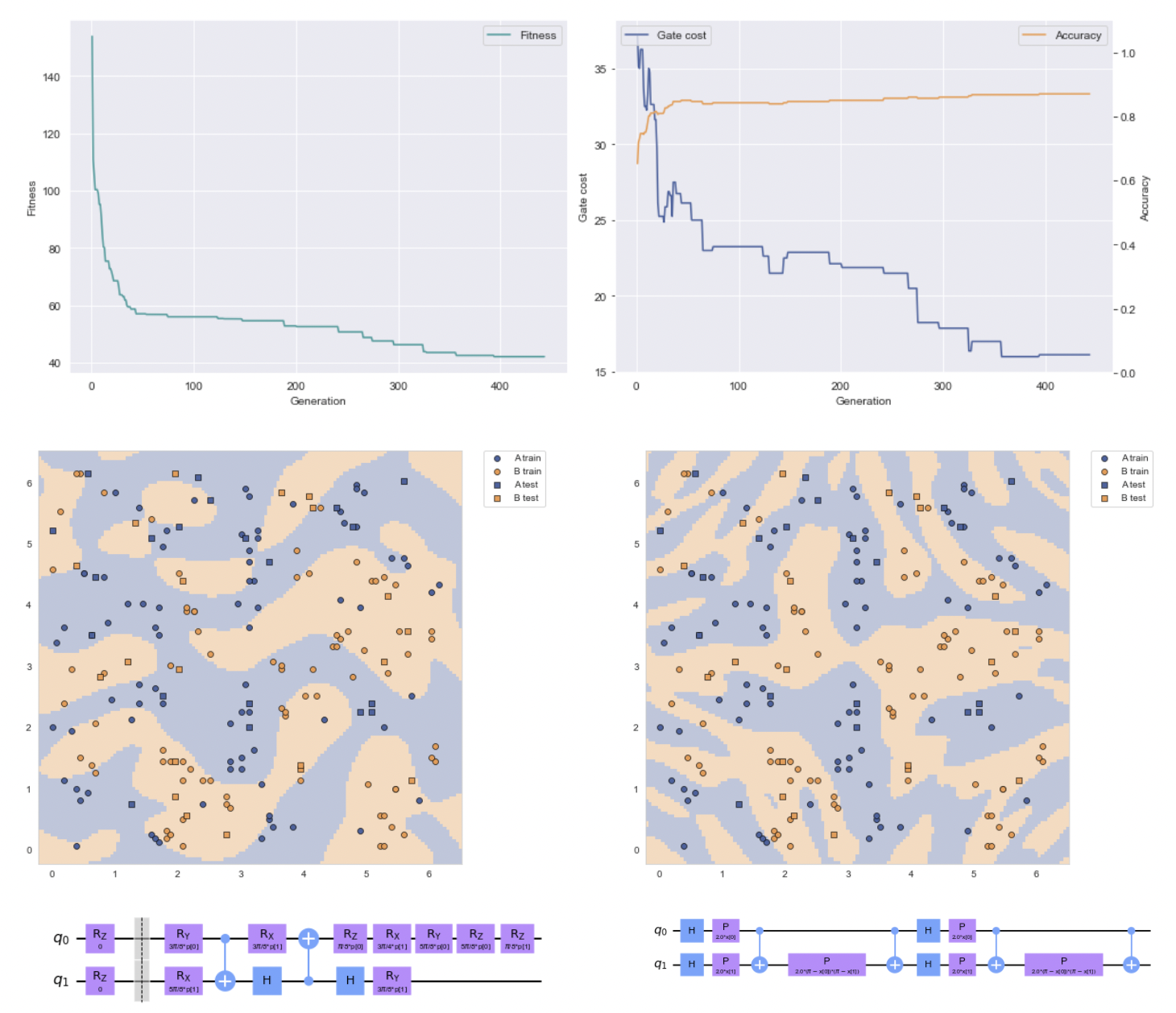}
\end{figure}

Secondly, we consider the ad hoc data generated by Qiskit \cite{ALE19}. 
For genetic algorithm, we choose the weight $w$ = 80 in equation \ref{GateCost} because the data here is more
complex, which we need to put more emphasize on the accuracy factor.
We obtain 90$\%$ accuracy and we compare the circuit with 
ZZFeatureMap \cite{HCT19}, though
ZZFeatureMap will always get perfect accuracy because the ad hoc dataset is generated
by ZZFeatureMap. We have an advantage by gate cost is 23 in compare with ZZFeatureMap's
gate cost is 34 as illustrated in Figure \ref{ga_adhoc}. From lots of experiments,
we observe that with high accuracy circuits, more gate cost circuit makes the
QSVM model more over fitting.

For variational circuits, hardware efficiency ansatz has produced an accuracy of 
92.5$\%$, while unitary decomposition ansatz has reached almost perfect accuracy
with 97.5$\%$. We choose to use four depths of hardware efficient ansatz in order
to have similar numbers of parameters with unitary decomposition ansatz, and it turns
out that with more complex datasets, unitary decomposition ansatz have better performance
in general. 

\begin{figure}[htbp]
  \centering
  \caption{Ad hoc dataset using hardware efficient ansatz.}
  \label{he_adhoc}
  \includegraphics[width=0.48\textwidth]{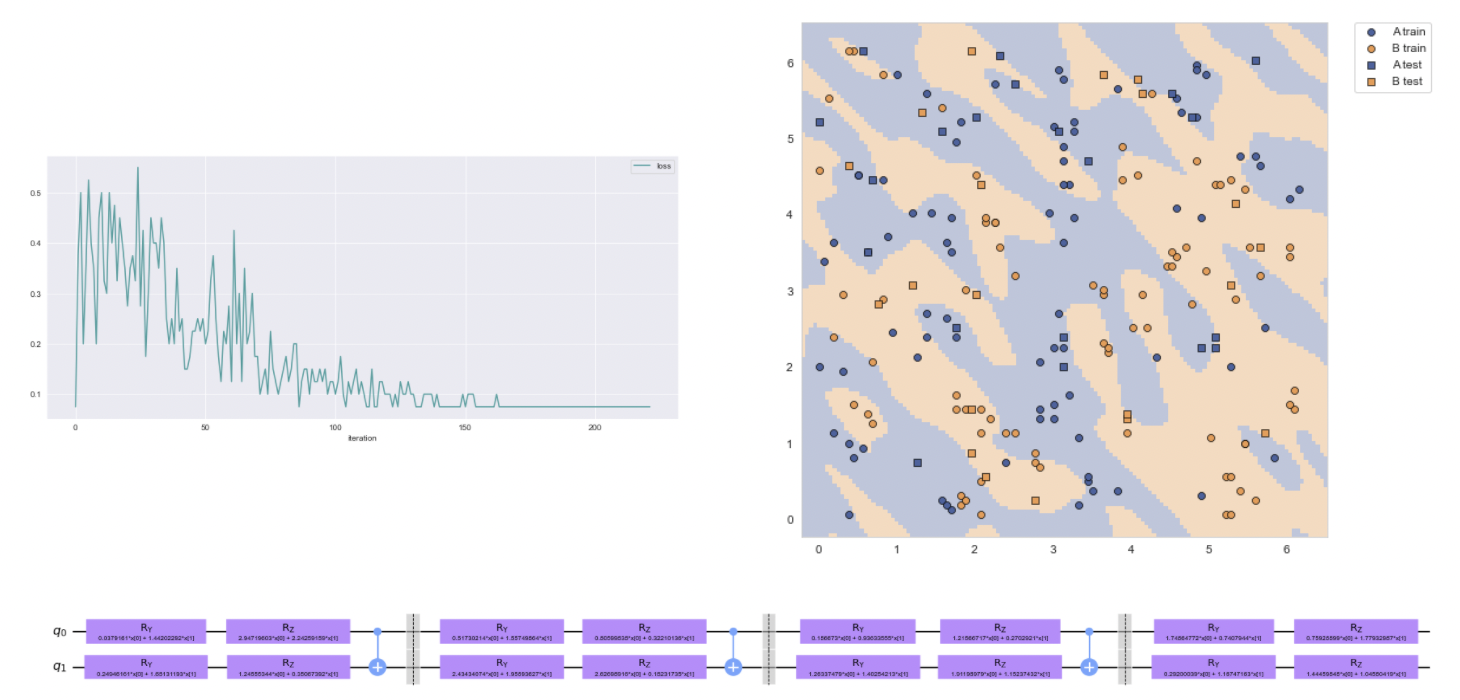}
\end{figure}

\begin{figure}[htbp]
  \centering
  \caption{Ad hoc dataset using unitary decomposition ansatz.}
  \label{ud_adhoc}
  \includegraphics[width=0.48\textwidth]{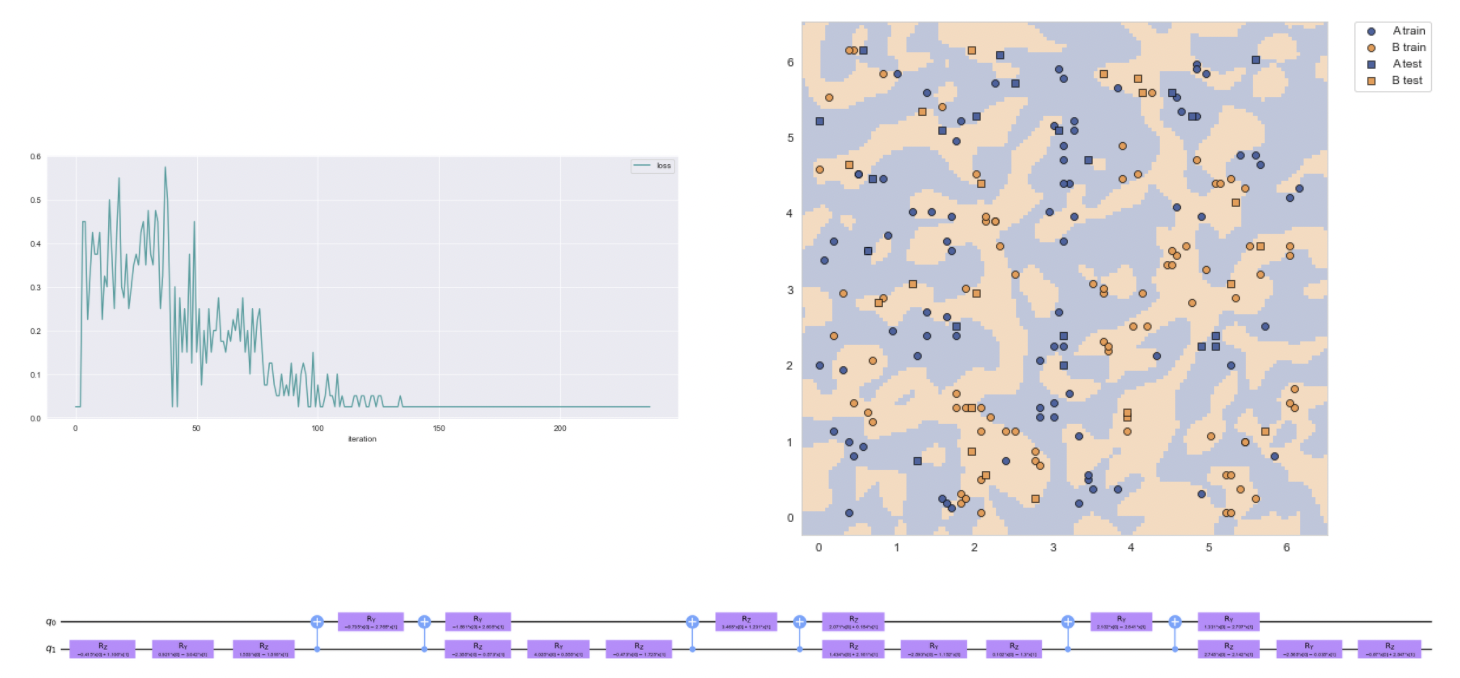}
\end{figure}

We compared different kernels and quantum feature map generating methods together, 
the RBF (Guassian) kernel can already obtain well performance with simple datasets
and using ZZFeatureMap for the quantum kernel is only suitable for the ad hoc dataset.
Comparing genetic algorithm (GA), hardware efficient ansatz (HE) and 
unitary decomposition ansatz (UD), genetic algorithm works well with simple 
datasets and can efficiently decrease the gate cost, but with more complex 
datasets, it is hard to find global solution and consumes a lot of time. 
Hardware efficient ansatz is flexible with gate cost by choosing suitable depths, 
and have slightly better accuracy than genetic algorithm. Unitary decomposition ansatz
has the best accuracy overall, but has the disadvantage of high fixed gate cost.

\begin{ruledtabular}
\begin{table}[htbp]
    \centering
    \caption{Comparing classification methods.}
    \label{table_compare}
    \begin{tabular}{lcccccc}
      \multirow{2}{*}{Method} & \multicolumn{3}{c}{Moonshape} & \multicolumn{3}{c}{Ad hoc}     \\
                              & acc      & gate   & time      & acc          & gate    & time  \\
      \\[-1.2em]
      \hline
      \\[-1.2em]
      SVM-linear              & 80.0$\%$ &                &   & 60.0$\%$     &              &  \\
    %   \\[-1em]
      SVM-RBF                 & 100$\%$  & $(\gamma=0.5)$ &   & 80.0$\%$     & $(\gamma=8)$ &  \\
    %   \\[-1em]
      QSVM-ZZ                 & 50.0$\%$ & 34             &   & 100$\%$      & 34           &  \\
      \\[-1.2em]
      \hline
      \\[-1.2em]
      QSVM-GA                 & 100$\%$  & {\bf3} & 66.30s    & 90.0$\%$     & {\bf23} & 2067.96s \\
    %   \\[-1em]
      QSVM-HE                 & 100$\%$  & 4      & 33.51s    & 92.5$\%$     & 31      & 197.68s  \\
    %   \\[-1em]
      QSVM-UD                 & 100$\%$  & 38     & 209.61s   & {\bf97.5$\%$}& 38      & 220.06s  \\
    \end{tabular}
\end{table}
\end{ruledtabular}

\subsection{Covariant quantum kernel dataset}\label{sec5-3}
Thirdly, we compare our genetic algorithm model
with quantum kernel training \cite{GGC22}, which is also designed to find a proper kernel
via trainable parameters. Because our dataset does not fit the group structure with the
circuit they designed, we try to compare by implementing our model to their dataset
\cite{ALE19} with 14 features generated by specific group cosets.

\begin{ruledtabular}
\begin{table}[htbp]
    \centering
    \caption{Comparison with covariant kernel.}
    \label{table_covariant}
    \begin{tabular}{ l  c c }
    %   \\[-1em]
      & Quantum covariant kernel & QSVM multi-fitness\\
    %   \\[-1em]
      \hline
    %   \\[-1em]
      Accuracy  & 100$\%$ & 100$\%$                \\
    %   \\[-1em]
      Gate cost & 51      & 41$\rightarrow${\bf 3} \\
    \end{tabular}
\end{table}
\end{ruledtabular}

\begin{figure}[htbp]
    \centering
    \caption{Comparison with Covariant kernel.}
    \label{covariant}
    \includegraphics[width=0.48\textwidth]{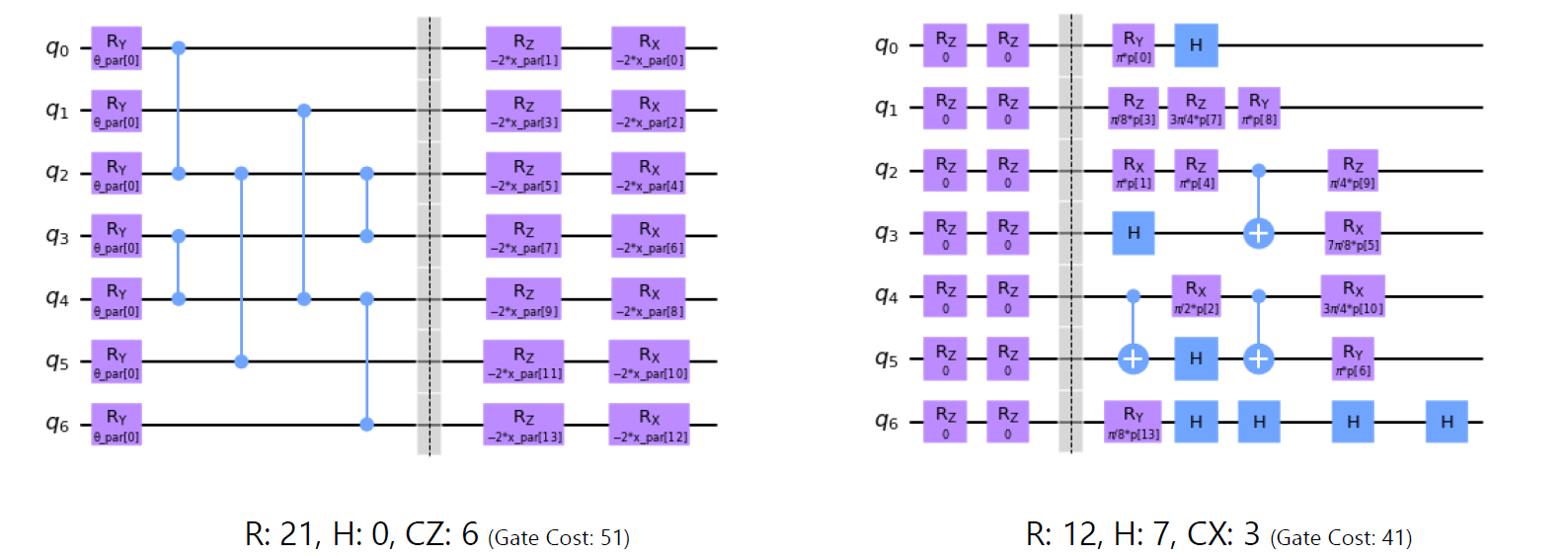}
\end{figure}

\begin{figure}[htbp]
    \centering
    \caption{Optimizing circuit.}
    \label{optimized}
    \includegraphics[width=0.48\textwidth]{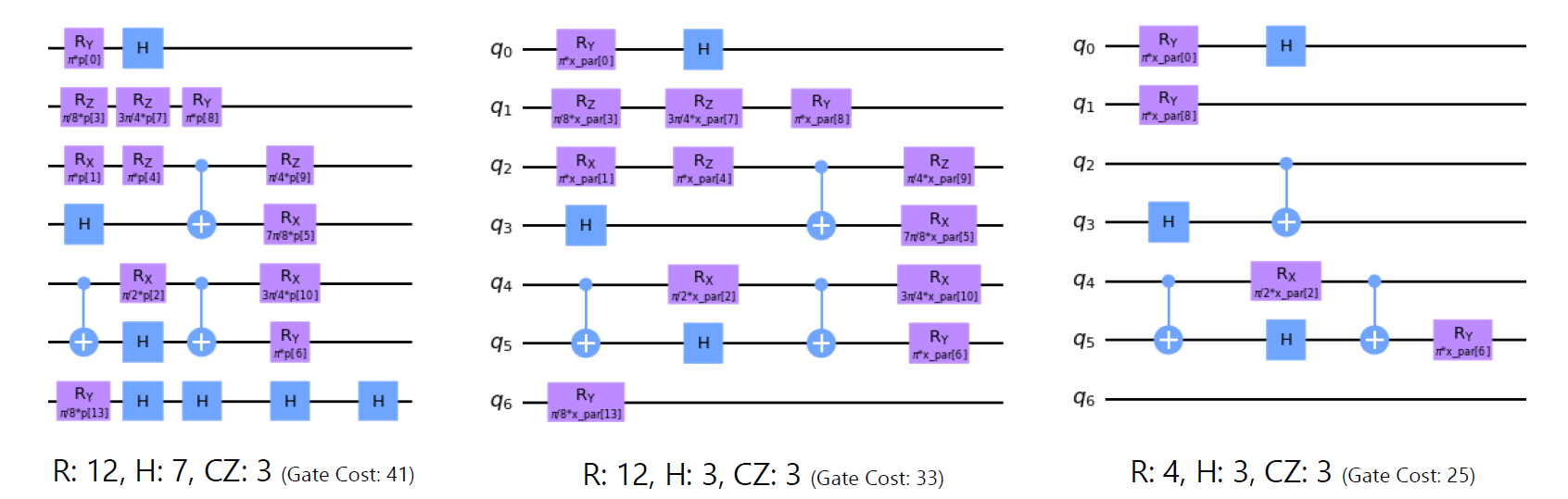}
    \includegraphics[width=0.48\textwidth]{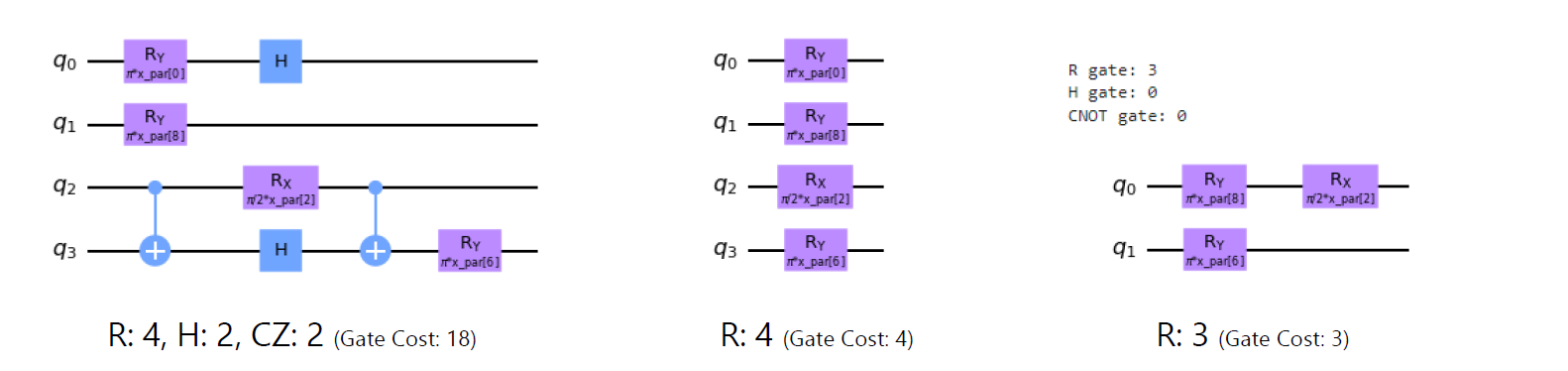}
\end{figure}

After experiments, we found a circuit with the gate cost of 41 and perfect accuracy,
the rotation gates on the left side of the barrier are identity gates and can be ignored.
It is obvious that
the 4 Hadamard gates at the last qubit does not effect and can be removed to decrease
the gate cost to 33. The circuit has already outperformed covariant quantum kernel's
circuit, due to the fact that their circuit is with the gate cost of 51 (21 R gates and
6 CZ gates). Moreover, the circuit can be optimized as shown in Figure \ref{optimized},
we test the necessity of each rotation gate in order to check
if every feature is useful, and it turns out that only 4 features are essential here.
Again, it is trivial to remove qubits without feature parameters involved, the last qubit
without any gates and the two qubits with only a Hadamard gate and CNOT gate can be removed.
We repeat the steps before, as we found that the two Hadamard gates and CNOT gates can also
be removed, that leaves 4 rotation gates. Again, we test the necessity of each rotation gate
where only 3 are essential, 2 among them can be added on the same qubit in order to decrease
the size of circuit.
As a result, we optimized the circuit to a simple uncorrelated
circuit with the gate cost of 3 and perfect classification accuracy.
More details are available at my Github \cite{CKmethod}. 

%---------------- Section 5------------------------
\section{Conclusion} \label{sec6}

In this work, we optimize the quantum feature map of QSVM using genetic algorithm
and quantum variational circuit. 
A quantum circuit is generated as a feature map
that maps the data points into Hilbert space, and then classified with quantum kernel
method applied QSVM. The genetic algorithm 
{modified in this study \cite{ARG21},}
aims to maximize the accuracy of QSVM while
minimizing the gate cost of quantum circuit with a multi-objective fitness function.
The variational circuit learns through paremeters with a given ansatz, based on 
special unitary groups and unitary matrix decomposition.
We test our models 
on three different datasets, a simpler moonshape dataset,
a more complex ad hoc dataset and a specifically generated dataset for comparison.
We all found high accuracy of classification, 
where more complex dataset is needed to increase the penalty weight $w$
to decrease the influence of gate cost or increase the circuit depth in order to 
have more learnable parameters.

There are many more things to explore, including testing the circuit by variational
quantum classifier (VQC), quantum neural network (QNN) or quantum distance-based
classifier. Adding trainable parameters $\theta_i$ to the circuit that can be
optimized using SPSA is also another strategy. We believe it is possible to find
suitable feature maps that classification can be well-performed in the Hilbert space.

%----------------------------------------------

\section{Acknowledgement}
Thank the authors of \cite{ARG21} for helpful suggestions and communications,
which help improve the presentation of this paper.

%---------------- References---------------------

% \bibliographystyle{apsrev4-2}


\begin{thebibliography}{99}

    \bibitem{ADB14}
    {E.B.\ Alaia, I.H.\ Dridi, H.\ Bouchriha and P.\ Borne},
    {\em Genetic algorithm with pareto front selection for multi-criteria optimization of
    multi-depots and multi- vehicle pickup and delivery problems with time windows},
    2014 15th International Conference on Sciences and Techniques of Automatic Control
    and Computer Engineering, pp. 488-493, (2014).
    
    \bibitem{ALE19}
    {G.\ Aleksandrowicz} et al,
    {\em Qiskit: An Open-source Framework for Quantum Computing},
    Zenodo, (2019).
    
    \bibitem{ARG21}
    {S.\ Altares-López, A.\ Ribeiro, J.J.\ García-Ripoll},
    {\em Automatic design of quantum feature maps},
    Quantum Science and Technology, vol. 6, no. 4, (2021).

    \bibitem{BWP17}
    {J.\ Biamonte, P.\ Wittek, N.\ Pancotti, P.\ Rebentrost, N.\ Wiebe and S.\ Lloyd},
    {\em Quantum machine learning},
    Nature 549, pp. 195-202, (2017).  

    \bibitem{BYH22}
    {A.\ Baughman, K.\ Yogaraj, R.\ Hebbar, S.\ Ghosh, R.U.\ Haq and Y.\ Chhabra},
    {\em Study of Feature Importance for Quantum Machine Learning Models},
    arXiv:2202.11204, (2022).
    
    \bibitem{CSU20}
    {D.\ Chivilikhin, A.\ Samarin, V.\ Ulyantsev, I.\ Iorsh, A.R.\ Oganov and O.\ Kyriienko},
    {\em MoG-VQE: Multiobjective genetic variational quantum eigensolver},
    arXiv:2007.04424, (2020).

    \bibitem{CYH16}
    {P.\ Chen, L.\ Yuan, Y.\ He and S.\ Luo},
    {\em An improved SVM classifier based on double chains quantum genetic
    algorithm and its application in analogue circuit diagnosis},
    Neurocomputing, vol. 211, pp. 202-211, (2016). 

    \bibitem{GA11}
    {M.\ Gönen and E.\ Alpaydin},
    {\em Multiple Kernel Learning Algorithms},
    Journal of Machine Learning Research, vol. 12, pp. 2211-2268, (2011).
    
    \bibitem{GGC22}
    {J.R.\ Glick, T.P.\ Gujarati, A.D.\ Córcoles, Y.\ Kim, A.\ Kandala, J.M. Gambetta and K.\ Temme},
    {\em Covariant quantum kernels for data with group structure},
    APS March Meeting 2022, vol. 67, no. 3, (2022).
    
    \bibitem{HCT19}
    {V.\ Havlíček, A.D.\ Córcoles, K.\ Temme, M.\ Takita,, M.\ Kandala,
    J.M.\ Chow and J.M.\ Gambetta},
    {\em Supervised learning with quantum-enhanced feature spaces},
    Nature 567, pp. 209-212, (2019).
    
    \bibitem{ICK16}
    {R.\ Iten, R.\ Colbeck, I.\ Kukuljan, J.\ Home, and M.\ Christandl},
    {\em Quantum circuits for isometries},
    Physical Review A 93, 032318, (2016).

    \bibitem{KMT17}
    {A.\ Kandala, A.\ Mezzacapo, K.\ Temme, A.W.\ Harrow, A.\ Brink,
    J.M.\ Chow and J.M.\ Gambetta},
    {\em Hardware-efficient variational quantum eigensolver for small molecules and quantum magnets},
    Nature 549, pp. 242-246, (2017).

    \bibitem{KSA21}
    {A.M.\ Krol, A.\ Sarkar, I.\ Ashraf, Z.\ Al-Ars and K.\ Bertels},
    {\em Efficient decomposition of unitary matrices in quantum circuit compilers},
    arXiv:2101.02993, (2021).

    \bibitem{LLK06}
    {S.\ Lee, S.J.\ Lee, T.\ Kim, J.S.\ Lee, J.\ Biamonte and M.\ Perkowski},
    {\em The cost of quantum gate primitives},
    Journal of Multiple-Valued Logic and Soft Computing, vol. 12, No. 5-6. pp. 561-573, (2006).
    
    \bibitem{LIN16}
    {H. Lingaraj},
    {\em A Study on Genetic Algorithm and its Applications},
    International Journal of Computer Sciences and Engineering, vol. 4, pp. 139-143, (2016).
    
    \bibitem{LAT21}
    {Y.\ Liu, S.\ Arunachalam and K.\ Temme},
    {\em A rigorous and robust quantum speed-up in supervised machine learning},
    Nature Physics, vol. 17, pp. 1013-1017, (2021).
    
    \bibitem{MS21}
    {M.\ Mafu and M.\ Senekane},
    {\em Design and Implementation of Efficient Quantum Support Vector Machine},
    2021 International Conference on Electrical, Computer and Energy Technologies, pp. 1-4, (2021).
    
    \bibitem{MBS22}
    {S.\ Moradi, C.\ Brandner, C.\ Spielvogel, D.\ Krajnc, S.\ Hillmich, R.\ Wille,
    W.\ Drexler and L.\ Papp},
    {\em Clinical data classification with noisy intermediate scale quantum computers},
    Scientific Reports, vol. 12, no. 1851, (2022).
    
    \bibitem{PMS14}
    {A.\ Peruzzo, J.\ McClean, P.\ Shadbolt, M.\ Yung, X.\ Zhou, A.\ Aspuru-Guzik and A.L.\ O'Brien},
    {\em A variational eigenvalue solver on a photonic quantum processor},
    Nature Communications, vol. 5, 4213  (2014).

    \bibitem{PQW20}
    {J.E.\ Park, B.\ Quanz, S.\ Wood, H.\ Higgins and R.\ Harishankar},
    {\em Practical application improvement to Quantum SVM: theory to practice},
    arXiv:2012.07725, (2020).
    
    \bibitem{PVG11}
    {F.\ Pedregosa, G.\ Varoquaux and A.\ Gramfort} et al,
    {\em Scikit-learn: Machine Learning in Python},
    Journal of Machine Learning Research,  vol. 12, pp. 2825-2830, (2011).

    \bibitem{RML14}
    {P.\ Rebentrost, M.\ Mohseni, and S.\ Lloyd},
    {\em Quantum Support Vector Machine for Big Data Classification},
    Physical Review Letters 113, 130503, (2014).

    \bibitem{SBM06}
    {V.V.\ Shende, S.S.\ Bullock, and I.L.\ Markov},
    {\em Synthesis of Quantum-Logic Circuits},
    IEEE Transactions on Computer-Aided Design of Integrated Circuits and Systems,
    vol. 25, no. 6, (2006).
    
    \bibitem{SCH21}
    {M.\ Schuld},
    {\em Quantum machine learning models are kernel methods},
    arXiv:2101.11020, (2021).
    
    \bibitem{SK19}
    {M.\ Schuld and N.\ Killoran},
    {\em Quantum Machine Learning in Feature Hilbert Spaces},
    Physical Review Letters 122, 040504, (2019).

    \bibitem{SMB04}
    {V.V.\ Shende, I.L.\ Markov, and S.S.\ Bullock},
    {\em Minimal universal two-qubit controlled-NOT-based circuits},
    Physical Review A 69, 062321, (2004).
    
    \bibitem{SSP14}
    {M.\ Schuld, I.\ Sinayskiy and F.\ Petruccione},
    {\em An introduction to quantum machine learning},
    Contemporary Physics, vol. 56, (2015).
    
    \bibitem{SYG20}
    {Y.\ Suzuki, H.\ Yano, Q.\ Gao, S.\ Uno, T.\ Tanaka, M.\ Akiyama and N.\ Yamamoto},
    {\em Analysis and synthesis of feature map for kernel-based quantum classifier},
    Quantum Machine Intelligence, vol. 2, no. 9, (2020).

    \bibitem{VW04}
    {F.\ Vatan and C.\ Williams},
    {\em Optimal quantum circuits for general two-qubit gates},
    Physical Review A 69, 032315, (2004).
  
    \bibitem{CKmethod}
    {B.S.\ Chen and J.\ Chern},
    {\em Generating quantum feature maps for SVM classifier},
    GitHub repository,
    \url{https://github.com/doggydoggy0101/Quantum_feature_maps}

\end{thebibliography}
\end{document}